\documentclass[letterpaper,  
              ]{jacow}
\makeatletter%
	\ifboolexpr{bool{xetex}}
	 {\renewcommand{\Gin@extensions}{.pdf,%
	                    .png,.jpg,.bmp,.pict,.tif,.psd,.mac,.sga,.tga,.gif,%
	                    .eps,.ps,%
	                    }}{}
\makeatother

\ifboolexpr{bool{xetex} or bool{luatex}} 
 {}                                      
 {\usepackage[utf8]{inputenc}}           

\usepackage[USenglish]{babel}			 

\usepackage[final]{pdfpages}
\usepackage{multirow}
\usepackage{ragged2e}

\ifboolexpr{bool{jacowbiblatex}}%
 {%
  \addbibresource{jacow-test.bib}
  \addbibresource{biblatex-examples.bib}
 }{}
\newcommand{\Hsh}{\ensuremath{H_{\mathrm{sh}}}}

\begin{document}

\title{SRF Theory Developments from the Center for Bright Beams\thanks{This work was
supported by the US National Science Foundation under Award OIA-1549132, the Center for Bright Beams.}}

\author{D. B. Liarte\thanks{dl778@cornell.edu},
T. Arias, D. L. Hall, M. Liepe, J. P. Sethna, N. Sitaraman, \\
Cornell University, Ithaca, NY, United States of America \\
A. Pack, M. K. Transtrum, \\
Brigham Young University, Provo, UT, USA}
	
\maketitle

\begin{abstract}
We present theoretical studies of SRF materials from the Center for Bright Beams.
First, we discuss the effects of disorder, inhomogeneities, and materials anisotropy
on the maximum parallel surface field that a superconductor can sustain in an SRF
cavity, using linear stability in conjunction with Ginzburg-Landau and Eilenberger
theory. We connect our disorder mediated vortex nucleation model to current
experimental developments of Nb$_3$Sn and other cavity materials. Second, we use
time-dependent Ginzburg-Landau simulations to explore the role of inhomogeneities
in nucleating vortices, and discuss the effects of trapped magnetic flux on the residual
resistance of weakly- pinned Nb$_3$Sn cavities. Third, we present first-principles
density-functional theory (DFT) calculations to uncover and characterize the key
fundamental materials processes underlying the growth of Nb$_3$Sn. Our calculations
give us key information about how, where, and when the observed tin-depleted
regions form. Based on this we plan to develop new coating protocols to mitigate the
formation of tin depleted regions.
\end{abstract}

\section{Introduction}
\begin{figure}[!htb]
\centering
\includegraphics[width=0.9\linewidth]{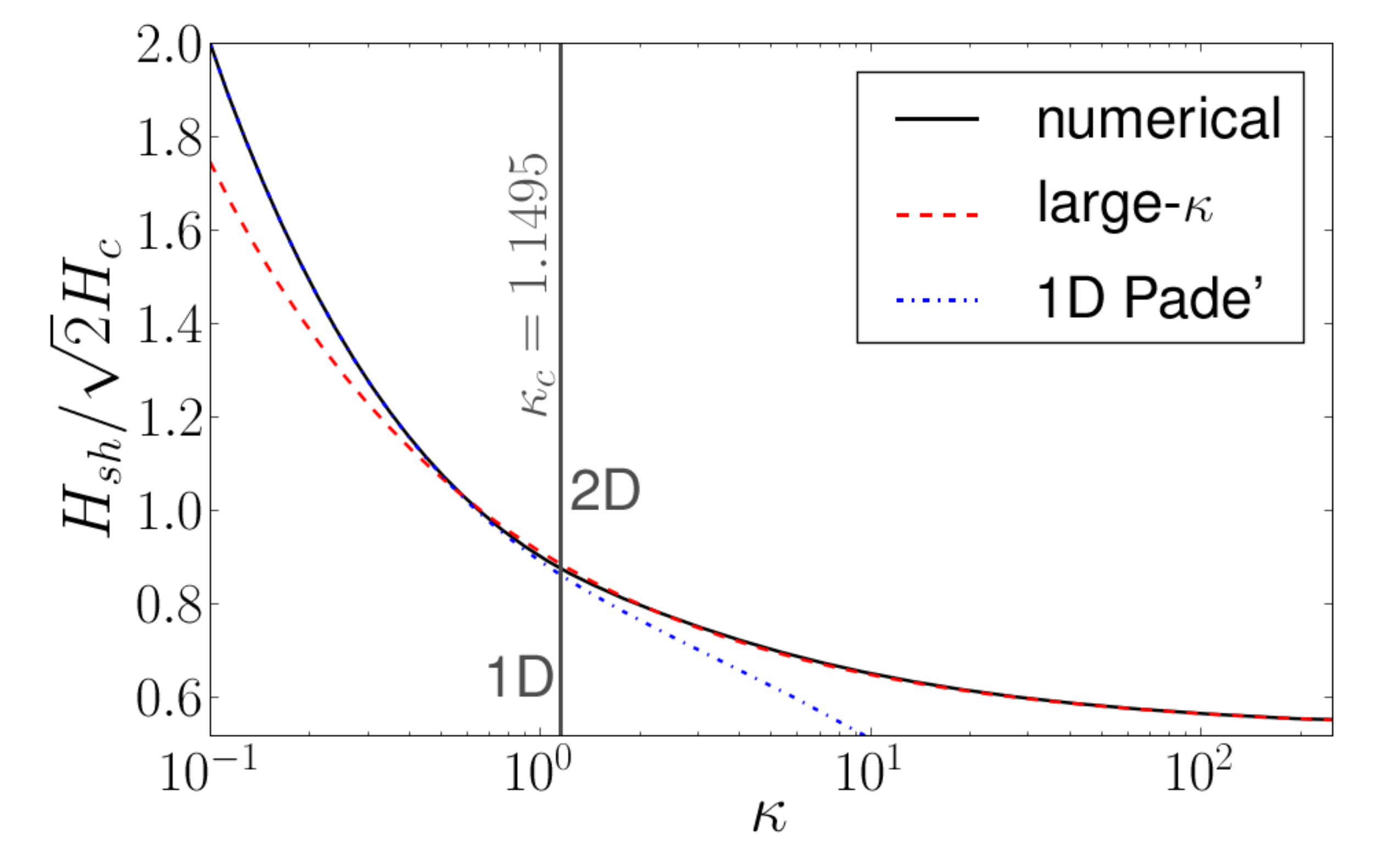}
\caption{From ref.~\cite{transtrum11}, showing a numerical estimate of \Hsh\ in
Ginzburg-Landau theory over many orders of magnitude of $\kappa$ (black solid
line), along with a large-$\kappa$ expansion (red dashed line), and a Pad\'e
approximation for small $\kappa$ (blue dotted-dashed line).}
\label{fig:HshVsKappa}
\end{figure}

The fundamental limit to the accelerating $E$-field in an SRF cavity 
is the ability of the superconductor to resist penetration of the associated
magnetic field $H$ (or equivalently $B$). SRF cavities are routinely run
at peak magnetic fields above the maximum field $H_{c1}$
sustainable in equilibrium; there is a metastable regime at higher fields
due to an energy barrier at the surface~\cite{bean64}. \Hsh\ marks the
stability threshold of the Meissner state. In Fig.~\ref{fig:HshVsKappa}
we show results from linear stability analysis~\cite{transtrum11}, valid near
$T_c$, for \Hsh\ as a function of the Ginzburg-Landau parameter $\kappa$,
the ratio $\lambda/\xi$ of the London penetration depth $\lambda$ to the
coherence length $\xi$. Niobium has $\kappa \approx 1.5$, most of the promising
new materials have large $\kappa$. At lower temperatures, one must move to 
more sophisticated Eliashberg theories~\cite{catelani08}, for which \Hsh\ is known
analytically for large $\kappa$; numerical studies at lower $\kappa$ are
in progress~\cite{catelaniUnpublished}. Broadly speaking, the results so far
for isotropic materials appear similar to those of Ginzburg-Landau.

This manuscript will briefly summarize theoretical work on \Hsh\ (the
threshold of vortex penetration and hence the quench field). 
First, we discuss the effect of materials anisotropy
on \Hsh~\cite{liarte16}. Second, we discuss theoretical
estimates of the effect of disorder~\cite{liarte17}, and
preliminary unpublished simulations of
the effects of surface roughness and materials inhomogeneity.
Third, we discuss key practical implications of theoretically calculated point
defect energies, interactions, relaxation times, and mobilities in the promising
new cavity material Nb$_3$Sn. Finally, some magnetic flux is trapped in
cavities during the cooldown phase, and the response of these flux lines
to the oscillating external fields appears to be the dominant source
of dissipation in modern cavities. We
model potentially important effects of multiple weak-pinning centers on this
dissipation due to trapped flux.

\section{The effect of materials anisotropy on the maximum field}
\label{sec:Anisotropy}

\begin{figure}[!htb]
\centering
\includegraphics[width=0.9\linewidth]{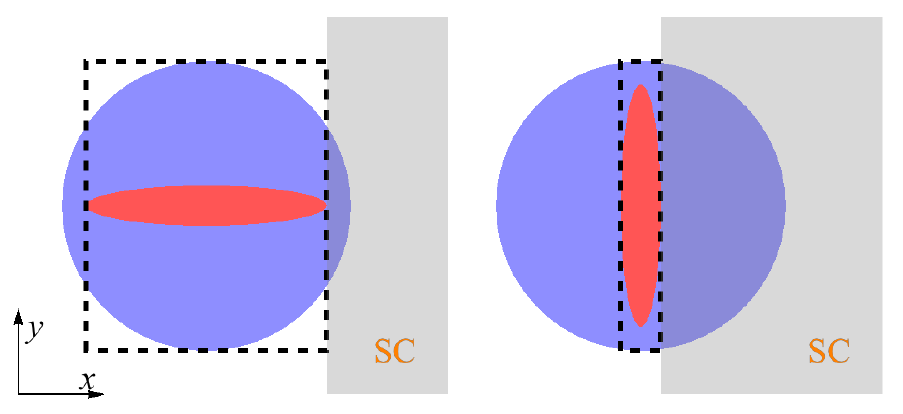}
\caption{From ref.~\cite{liarte17}, showing vortex (blue disk) and vortex core (red disk)
of zero-temperature MgB$_2$ in the $ac$ plane, with the external magnetic field parallel
to the normal of the plane of the figure. We have drawn the core region about 30 times
larger with respect to the penetration depth, so that the core becomes discernible.}
\label{fig:anisotropicVortex}
\end{figure}

Some of the promising new materials are layered, with strongly anisotropic
superconducting properties (MgB$_2$ and the pnictides, for example, but not
Nb$_3$Sn or NbN). Fig.~\ref{fig:anisotropicVortex} illustrates an anisotropic vortex
(magnetized region blue, vortex core red) penetrating into the surface
of a superconductor (grey). The anisotropy here is characteristic of MgB$_2$
at low temperatures, except that the vortex core is expanded by a factor
of 30 to make it visible. 

\begin{figure}[!htb]
\centering
\includegraphics[width=0.9\linewidth]{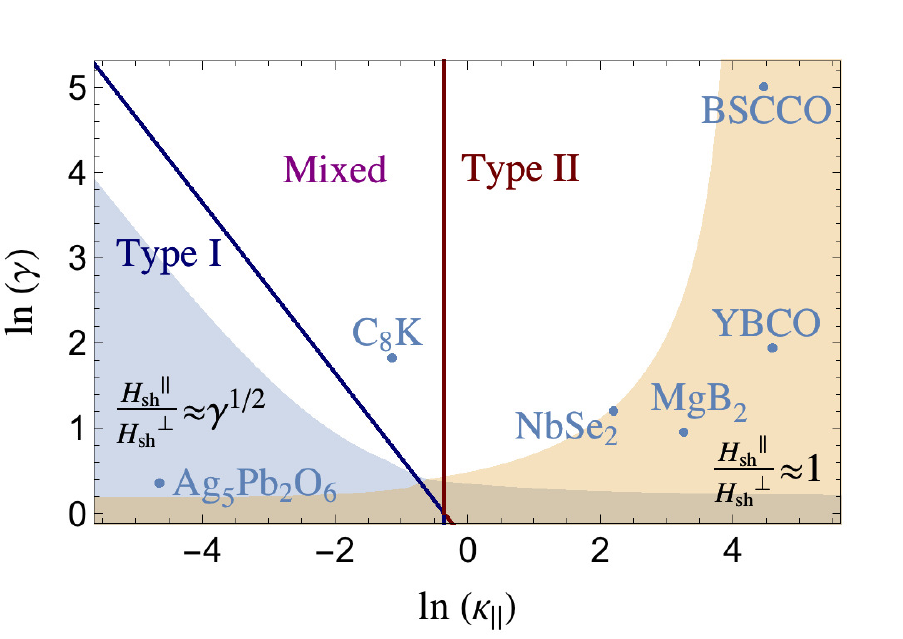}
\caption{From ref.~\cite{liarte16}, showing the phase diagram of anisotropic
superconductors in terms of mass anisotropy and GL parameters.
The shaded blue and orange regions correspond to regions where the
superheating field anisotropy can be approximated by $\gamma^{1/2}$ and
$1$, respectively, within 10\% of accuracy. Note that
the superheating field of MgB$_2$ is nearly isotropic near $T = T_c$. 
}
\label{fig:materialsAnisotropy}
\end{figure}

Near $T_c$, we find in ref.~\cite{liarte16} that a simple coordinate change
and rescaling maps the anisotropic system onto the isotropic case 
(Fig.~\ref{fig:HshVsKappa} above, as studied in ref.~\cite{transtrum11}).
We find, near $T_c$ where 
Ginzburg-Landau theory is valid, that \Hsh\ is nearly 
isotropic for large~$\kappa$ materials (Fig.~\ref{fig:materialsAnisotropy}.
At lower temperatures, different heuristic estimates of the effects of
anisotropy on \Hsh\ yield conflicting results. Further work at lower
temperatures could provide valuable insight into the possible role of
controlled surface orientation for cavities grown from these new materials.

\section{Disorder-mediated flux entry and materials anisotropy}

Defect regions and inhomogeneity of superconductor properties can
weaken the performance of SRF cavities. In ref.~\cite{liarte16} we used
simple estimates based on Bean and Livingston's energy barrier
arguments~\cite{bean64}, to estimate the effects of disorder in lowering
\Hsh\ by providing flaws that lower the barrier to vortex penetration.
Here we use these calculations to shed light about the relationship between tin
depleted regions, low critical temperature profiles, defect sizes and
quench fields.

Consider an external magnetic field $B$, parallel to the
surface of a semi-infinite superconductor occupying the half-space
$x>0$. If $B$ is larger than the lower critical field $B_{c1}$ (and
smaller than $B_{c2}$), the vortex lattice phase is thermodynamically
favored. However, if the field is not large enough, a newborn vortex
line near the superconductor surface will have to surpass an energy
barrier to penetrate the superconductor towards the bulk of the
material. This instability typically is surmounted by the simultaneous
entry of an entire array of vortices, whose interactions lower one another's
barriers. Disorder, in contrast, will lead to a localized region allowing
one vortex entry at a time. Bean and Livingston provided simple analytical
calculations for the energy barrier felt by one vortex line; we extended
their calculation to estimate the dirt needed to reduce this barrier to 
zero at a quench field $H_q < \Hsh$. 

\begin{figure}[!htb]
\centering
\includegraphics[width=0.9\linewidth]{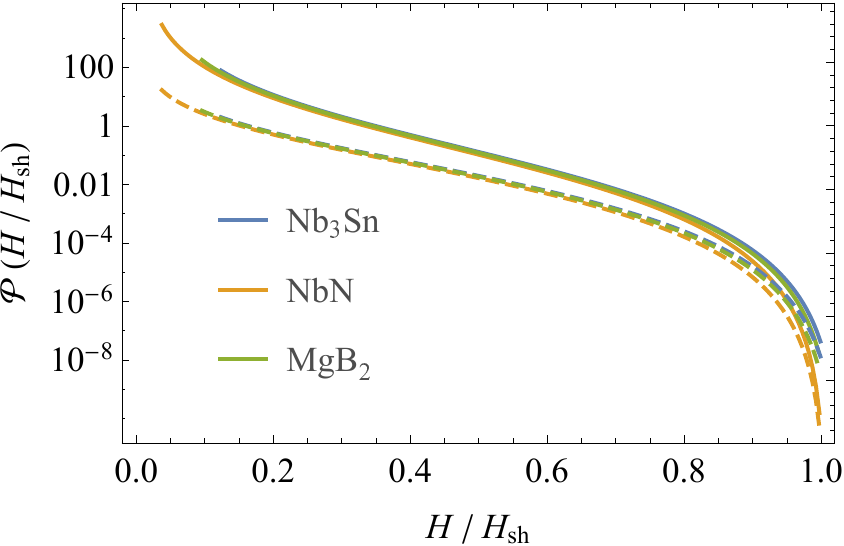}
\caption{From ref.~\cite{liarte17}, showing the reliability of vortex nucleation,
in a simple model of Gaussian random disorder, for three
candidate superconductors. Solid curves are for a 3D semicircular vortex
barrier model; dashed curves are for 2D pancake vortex nucleation in a
2D superconducting layer.}
\label{fig:reliabilityDirt}
\end{figure}

The new materials have larger $\kappa$, and in particular smaller vortex
core sizes $\xi$; naively one would expect vortex penetration when flaws
of size $\xi$ arise. Are these new materials far more sensitive to dirt
than niobium? Reassuringly, Fig.~\ref{fig:reliabilityDirt} shows that the low values
of the coherence length do not make these new materials substantially more 
susceptible to disorder-induced vortex penetration~\cite{liarte17}.

\begin{figure}[!htb]
\centering
(a)

\vspace{-0.21cm}
\includegraphics[width=0.9\linewidth]{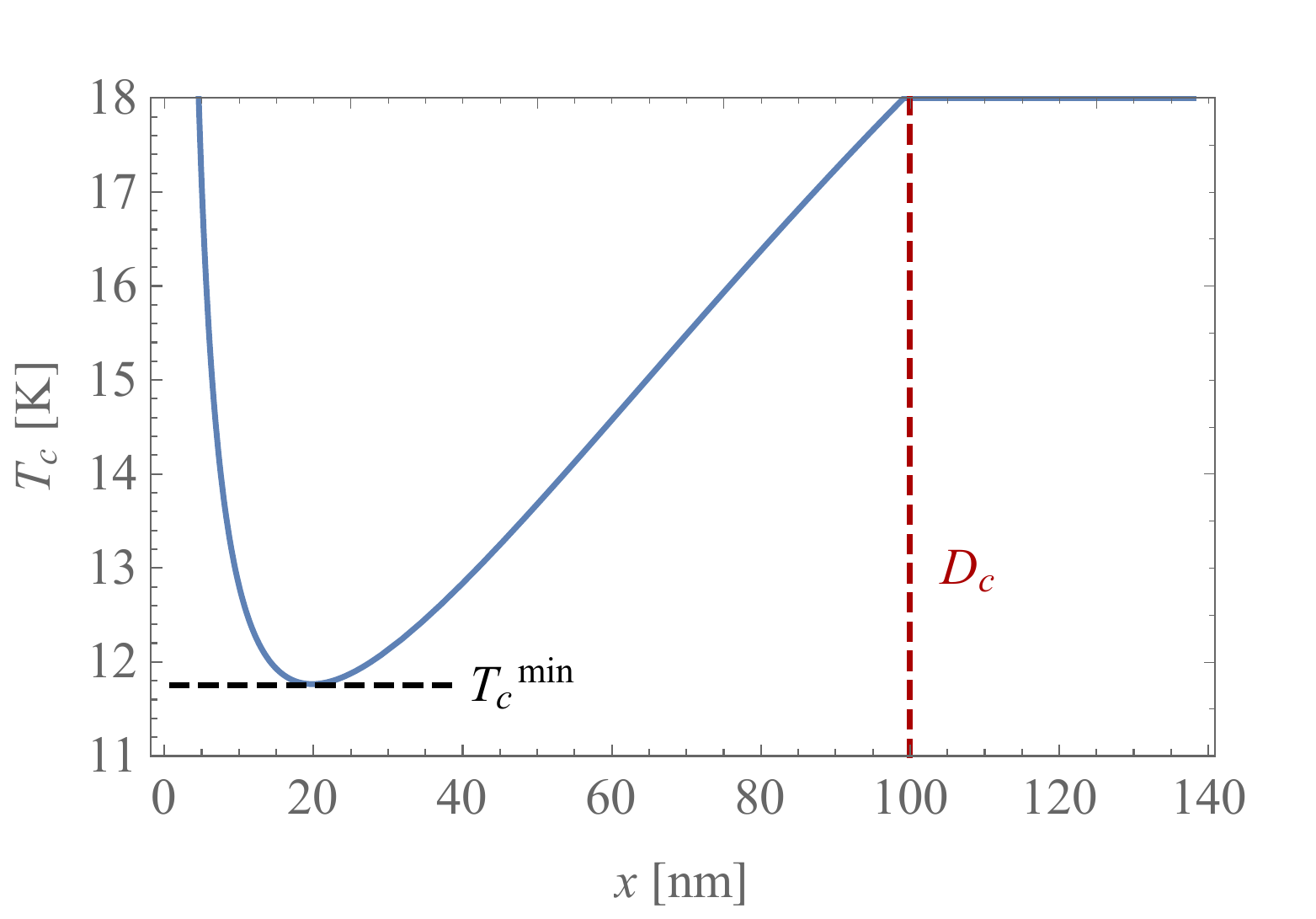}

(b)

\vspace{-1cm}
\includegraphics[width=0.9\linewidth]{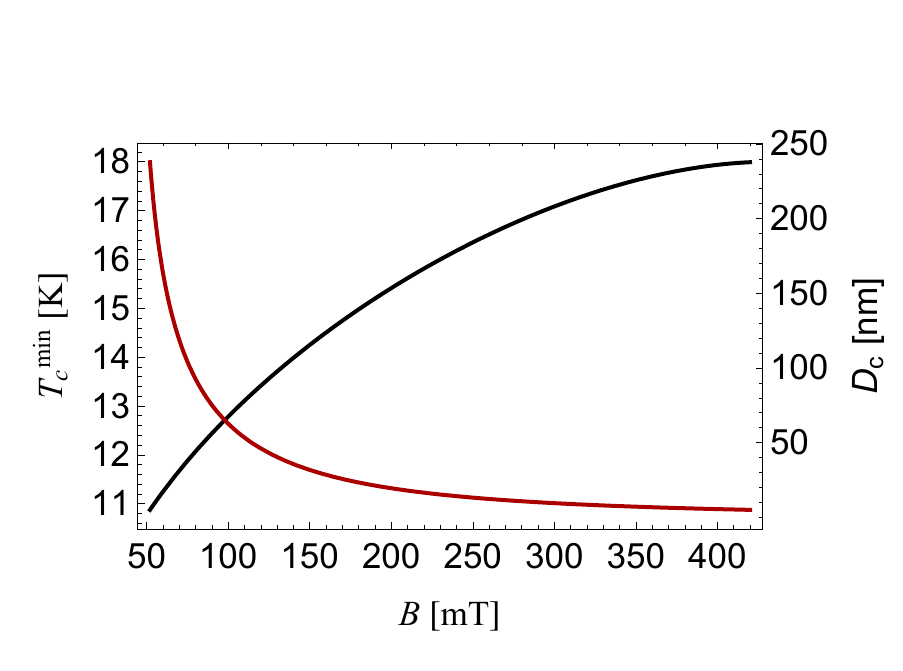}
\caption{(a) Critical temperature profile that allow nucleation of vortices
in Nb$_3$Sn cavities at a field of $\sim 77$mT. (b) Suppressed
superconducting transition temperature $T_c^{\min}$ (black), and flaw depth
$D_c$ (red), as a function of the quench field.}
\label{fig:quenchFieldPlot}
\end{figure}

We can use our model to estimate the suppressed superconducting
transition temperature $T_c^{\min}$ and the flaw depth $D_c$ needed
to allow vortex penetration, as a function of $H_q$ (or, in Tesla, $B_q$)
(Fig.~\ref{fig:quenchFieldPlot}). For Nb$_3$Sn we find a flaw size of
$D_c\sim 100$nm and $T_c^{min} \sim 12$K can allow vortex nucleation
and quenches at  fields of $\sim 77$mT (Fig.~\ref{fig:quenchFieldPlot}),
consistent with experimental results~\cite{hallIPAC17a}.

\section{Time-dependent Ginzburg-Landau simulations of rough surfaces
and disorder}

\begin{figure}[!htb]
\centering
\includegraphics[width=\linewidth]{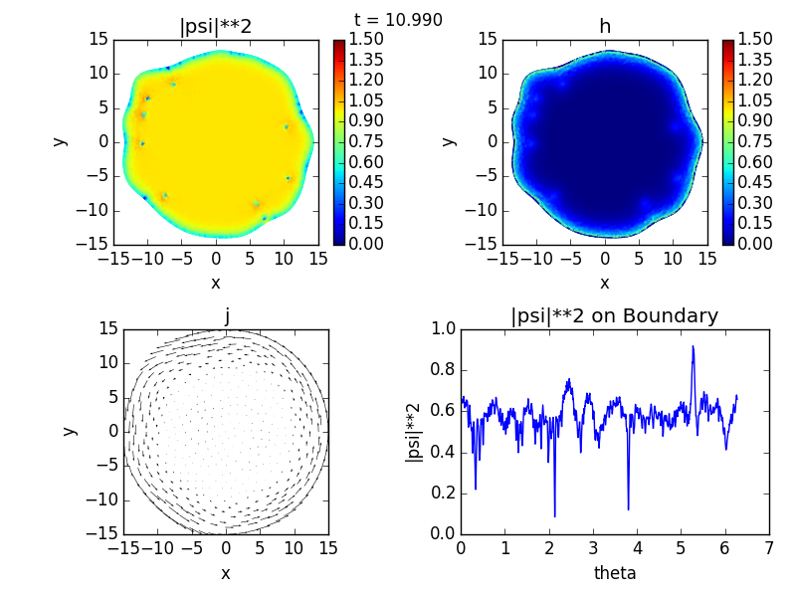}
\caption{Spatial dependence of the density of superconducting electrons
(top left), supercurrent (bottom left), and the induced magnetic field (top right). On
the bottom right, we show the variation of the order parameter around the perimeter
of the superconductor.}
\label{fig:surfaceRoughness}
\end{figure}

To quantify the dependence of \Hsh\ on surface roughness and disorder, 
we have developed a time-dependent Ginzburg-Landau simulation. 
Fig.~\ref{fig:surfaceRoughness} shows the density $|\psi|^2$ of superconducting
electrons at a field just above \Hsh (top left), showing the entry of several vortices
for a 2D system with an irregular surface. On the bottom left, we show the
corresponding supercurrent ${\mathbf{j}}$; on the top right we show the magnetic field
$H$ (perpendicular to the plane of the simulation), and on the bottom right we
show the effect of surface roughness on $|\psi(\theta)|^2$ around the 
perimeter. Our initial results quantify how inward-curving regions in the
plane perpendicular to the applied field on the perimeter can act as vortex
nucleation sites in this geometry.  An open question remains what the effect
of curvature and surface roughness have when oriented parallel to the applied
field.

The effect of roughness in Fig.~\ref{fig:surfaceRoughness} is to lower \Hsh\ by
a few percent.  By systematically varying the details of the roughness parameters,
we can use this tool to identify at what scale roughness will have significant impact on
vortex nucleation.  SRF cavity roughness can be smoothed to varying degrees.
Our TDGL environment can be used to find dangerous regimes or configurations
that can have serious consequences for cavity performance.

We can also use this tool as a way to explore vortex dynamics and the
effects of pinning sites on trapped residual magnetic flux. Pinning sites originate
from inhomogeneities in the material, such as grain boundaries or spatial
inhomogeneities in the alloy stoichiometry.  By incorporating this information
into our TDGL environment we can try to better understand the mechanisms
driving residual resistance for typical cavities.

\section{DFT calculations}

Nb$_3$Sn cavities are created by depositing tin vapor on the surface of a
niobium cavity, which reacts with the niobium to form an irregular surface layer of the
compound. Of interest are regions of ``tin depleted'' Nb$_3$Sn, known
to have a lower superconducting transition temperature than the surrounding Nb$_3$Sn.
These regions may be the nucleation centers responsible for quenches observed well
below \Hsh\ expected for perfect Nb$_3$Sn~\cite{hallIPAC17a}.

Density functional theory (DFT) can be used to study layer growth, tin depletion,
and other features of Nb$_3$Sn layers at the single-particle level. This information,
combined with experimental data and accounting for the effects of grain boundaries and
strain, makes it possible to build a multiscale model of layer growth.
\begin{figure}[!htb]
\centering
\includegraphics[width=0.6\linewidth]{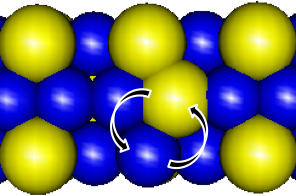}
\caption{Illustration of antisite disorder. We estimate that on the order of 1\% of
lattice sites are affected by antisite defects ``frozen in'' from the high coating temperature.
This would make them by far the most common point defect in Nb$_3$Sn layers.}
\label{fig:antisiteDisorder}
\end{figure}

\begin{figure}[!htb]
\centering
\includegraphics[width=\linewidth]{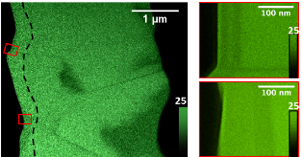}
\caption{Experimental data showing tin depletion. At left is a tin density map of a layer
cross section showing regions of significant (7-8\%) tin depletion, in this case mostly
deep in the layer relative to the RF penetration depth (dashed line). At right are close ups
showing slight (1-2\%) tin depletion right at the surface of the layer. Images by Thomas
Proslier at Argonne National Lab, received via personal communication with Daniel Hall.}
\label{fig:tinDepletion}
\end{figure}

Our initial work uses in-house DFT software to calculate defect formation and interaction
energies, impurity energies, and energy barriers in Nb$_3$Sn. We have found that
antisite disorder (Figure~\ref{fig:antisiteDisorder}), rather than impurities or vacancies,
likely sets the electron mean free path in Nb$_3$Sn and may also be responsible for
collective weak pinning. We have also found that under certain conditions during
growth, it is energetically favorable for Nb$_3$Sn to form at tin-depleted stoichiometry,
while during annealing existing Nb$_3$Sn near the surface or grain boundaries can
become tin-depleted by diffusion (Figure~\ref{fig:tinDepletion}). Either or both of these
tin depletion mechanisms may result in quench nucleation centers; by understanding
them we can for the first time make informed modifications to the coating process in an
attempt to limit tin depletion and produce better cavities.

\section{Dynamics of trapped vortices; potential role of weak pinning}

When the field is high enough for penetration of new vortices, one expects a 
cascade of vortices leading to a quench. Vortices trapped during the cooling
process, while not immediately fatal, do act as sources of residual resistance.
Experiments show that the non-BCS surface resistance is proportional to
the trapped flux, both for nitrogen-doped Nb cavities~\cite{gonnella16}
and for Nb$_3$Sn~\cite{hallIPAC17b}. This suggests that trapped vortices may be 
a dominant contribution to the quality factor of the cavity.

Previous studies of the residual resistance due to a trapped flux
line~\cite{gurevich13} focused on the Bardeen-Stephen viscous
dissipation~\cite{bardeen65} of a free line pinned a distance 
below the surface, as the external field drags the line through a
otherwise uniform superconducting medium. Experimental measurements
in nitrogen-doped Nb cavities showed good agreement to this theory,
except that the distance to the pinning center was presumed to change
linearly with the mean-free path~\cite{gonnella16} as it changes due to 
nitrogen doping. Since nitrogen (or other contaminant gases~\cite{koufalis17a,koufalis17b}) 
should act as weak pinning centers (with many impurities per coherence
length cubed), we have been modeling the role of weak pinning in 
vortex dissipation.
\begin{figure}[!htb]
\centering
\includegraphics[width=0.9\linewidth]{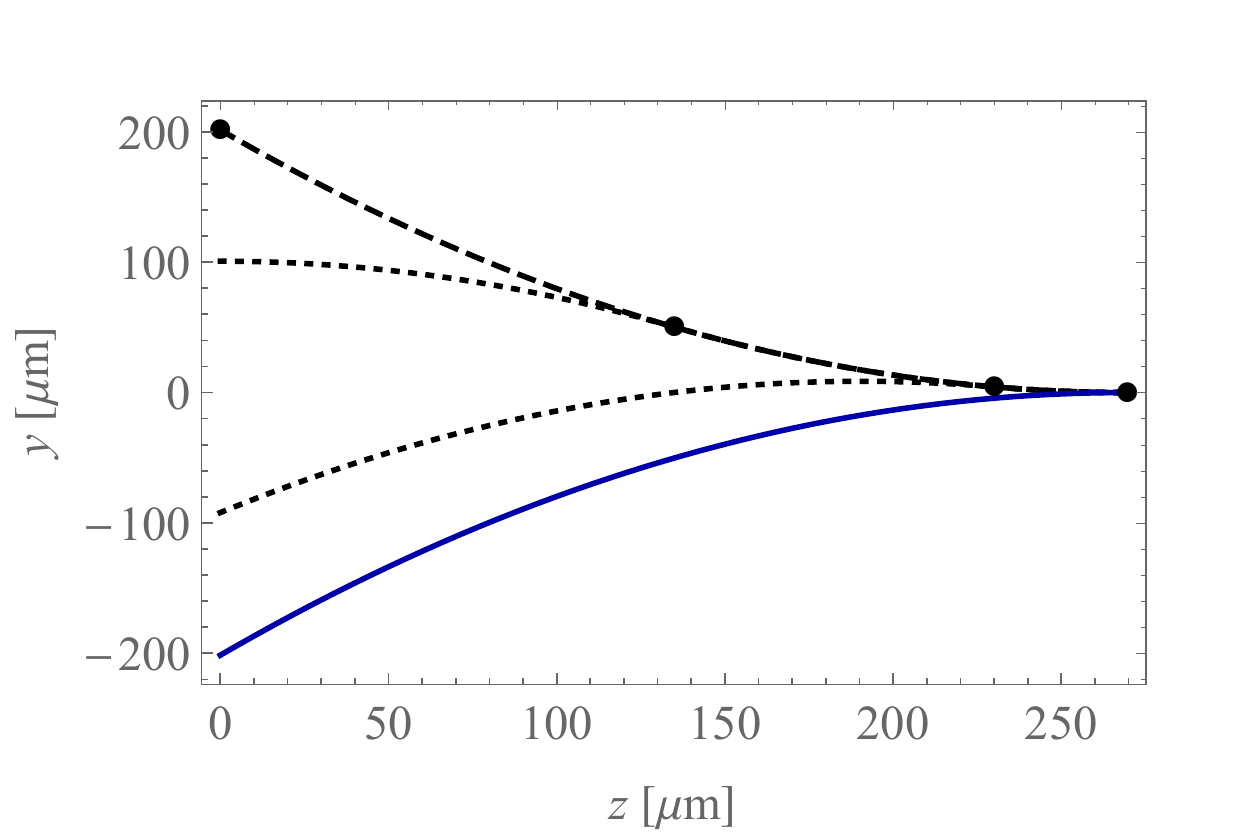}
\caption{From ref.~\cite{hallIPAC17b}, showing vortex line solutions at several
times, using measured parameters for Nb$_3$Sn.}
\label{fig:weakPinningSolution}
\end{figure}

\begin{figure}[!htb]
\centering
\includegraphics[width=0.9\linewidth]{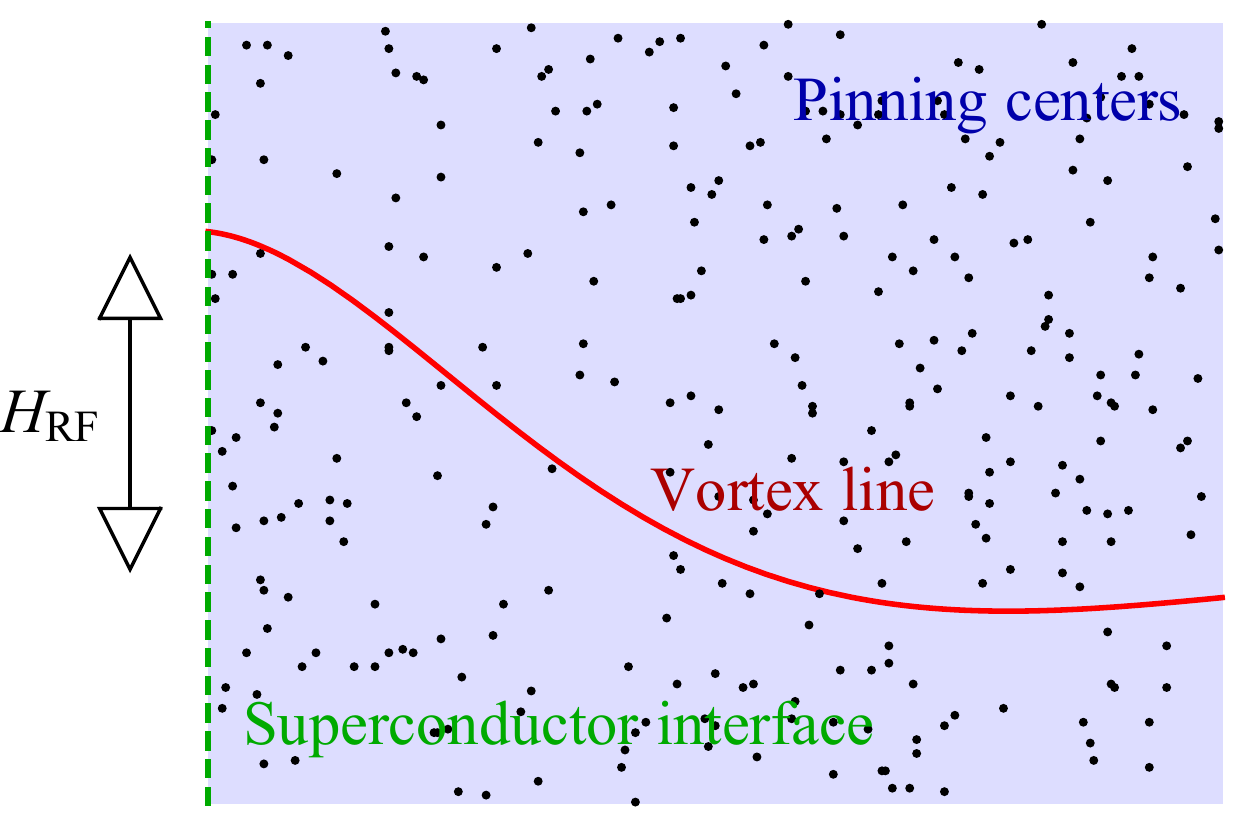}
\caption{Illustration of a vortex line (red curve) subject to an external rf magnetic field
and the collective action of many pinning centers.}
\label{fig:pinningEnvironment}
\end{figure}

Line defects pulled through a disordered medium is one of the classical
depinning transitions~\cite{fisher98}. The disorder acts as
a random potential, and macroscopically there is a threshold force per
unit length $f_{pin}$ needed to depin the line and start motion
(Fig.~\ref{fig:weakPinningSolution}). This depinning
transition is preceded by avalanches of all sizes (local regions of
vortex motion) and followed by fluctuations on all scales (jerky motion
of the vortex line in space and time). For our initial estimates, we
have ignored these fluctuations, using a `mean-field' model where our
superconducting vortex line has a threshold supercurrent
$j_d \propto f_{pin}^{2/3}$ for motion. We presume also that the 
energy dissipated is $f_{pin}$ times the area swept out by the vortex
as the external surface field pulls it to and fro (Fig.~\ref{fig:pinningEnvironment}).

\begin{figure}[!htb]
\centering
\includegraphics[width=0.9\linewidth]{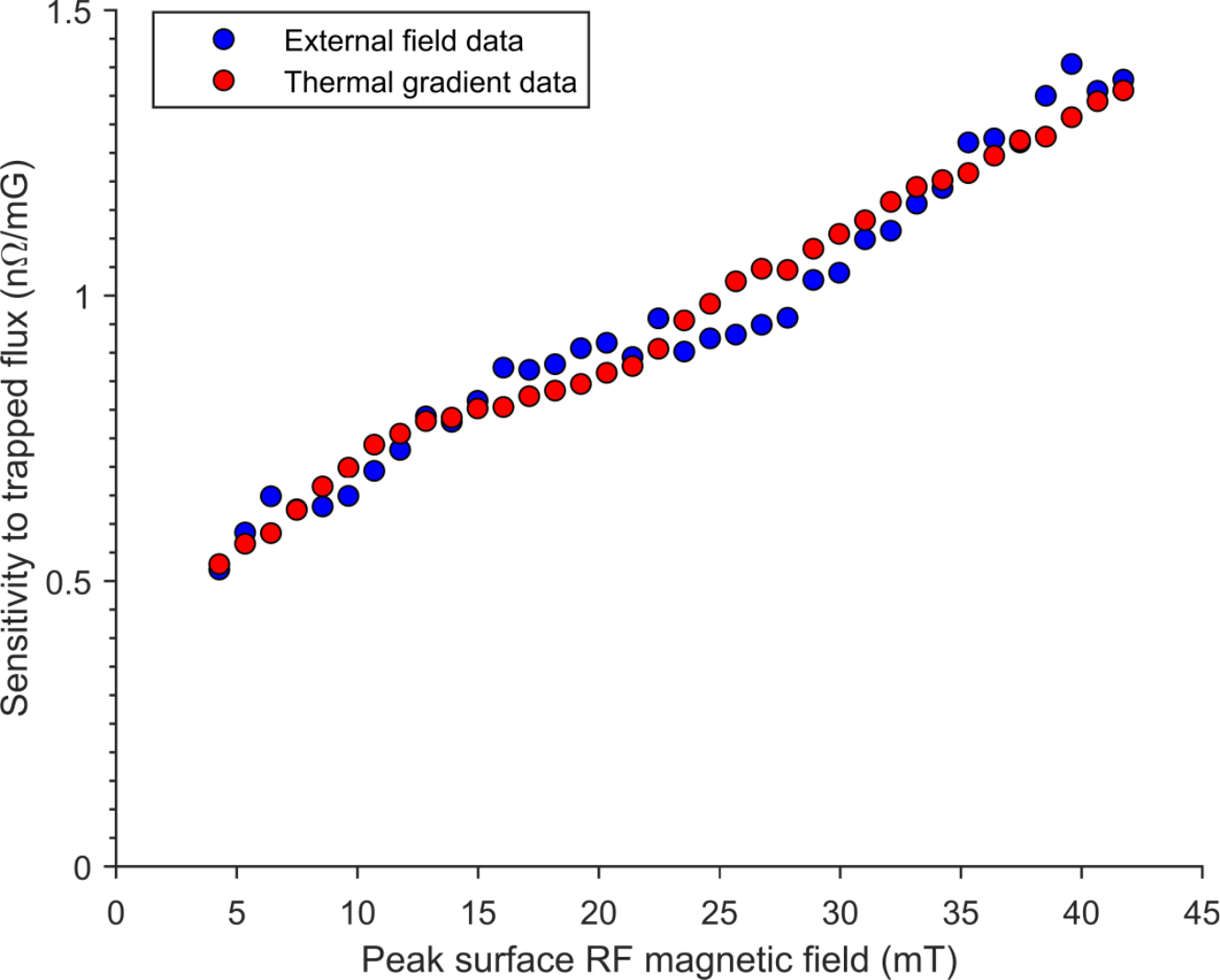}
\caption{From ref.~\cite{hallIPAC17b}, showing the sensitivity of residual
resistance to trapped magnetic flux, as a function of the peak rf field.}
\label{fig:sensitivity}
\end{figure}

The residual resistance measured in Nb$_3$Sn cavities shows a linear dependence
on the peak RF field (Fig.~\ref{fig:sensitivity}, \cite{hallIPAC17b}). The scaling
properties of the terms included in the earlier work~\cite{gurevich13}
all predict no dependence on the strength of the external oscillating field.
Our theory including weak pinning but ignoring the viscous dissipation
produces a dissipation that is linear in this external field. Our estimates,
however, suggest that our theory should be valid at MHz frequencies, but
at the operating GHz frequencies the viscous term must be important for
the energy dissipation. Our preliminary calculations suggest that incorporating
both can provide a reasonable explanation of the experimental data, but
we still do not obtain quantitative agreement.

\section{Conclusion}

The collaboration between scientists inside and outside traditional
accelerator physicists made possible by the Center for Bright Beams
has been immensely fruitful. This proceedings illustrates the richness
of the science at the intersection of accelerator experimentalists 
working on SRF cavities with condensed-matter physicists with interests
in continuum field theories and {\em ab-initio} electronic structure
calculations of materials properties. (One must also note the important
contributions of experimental condensed matter physicists in the collaboration.)
Current SRF cavities are pushing fundamental limits of superconductors,
and are a source of fascinating challenges for theoretical condensed-matter
physics. Conversely, we find that theoretical calculations are remarkably
fruitful in guiding and interpreting experimental findings.

\section{acknowledgment}
We thank Alex Gurevich for useful conversations.

\iffalse  
	\newpage
	\printbibliography

\else



\end{document}